\documentclass[11pt]{article}

\usepackage{subcaption,booktabs}

\usepackage{authblk}
\usepackage[english]{babel}
\usepackage[utf8]{inputenc}
\usepackage[T1]{fontenc}
\usepackage{amsmath,amsfonts,amsthm,amssymb}
\usepackage{mathtools}
\usepackage{graphicx}
\usepackage[table]{xcolor}
\usepackage{multirow}
\usepackage{xurl}
\usepackage{hyperref}
\usepackage{fullpage}
\usepackage[numbers]{natbib}
\hypersetup{
    colorlinks=true,
    linkcolor=red,
    filecolor=magenta,      
    urlcolor=blue,
    linkbordercolor = white
}

\usepackage{paralist}

\usepackage{enumitem}

\usepackage{algorithm2e}
\RestyleAlgo{ruled}
\SetKwComment{Comment}{/* }{ */}

\title{On Using Proportional Representation Methods as Alternatives to Pro-Rata Based Order Matching Algorithms in Stock Exchanges}
\author[1]{Sanjay Bhattacherjee}
\affil[1]{
	Institute of Cyber Security for Society and School of Computing \authorcr 
	Keynes College, University of Kent \authorcr 
	CT2 7NP, United Kingdom \authorcr 
	email: s.bhattacherjee@kent.ac.uk \authorcr
	ORCID: 0000-0002-3367-6192}
\author[2]{Palash Sarkar\thanks{Corresponding author.}}
\affil[2]{
	Indian Statistical Institute \authorcr 
	203, B.T. Road, Kolkata \authorcr
	India 700108 \authorcr 
	email: palash@isical.ac.in \authorcr
	ORCID: 0000-0002-5346-2650}

\date{\today}

\begin{document}

\maketitle

\begin{abstract}
	The first observation of the paper is that methods for determining proportional representation in electoral systems may be suitable as alternatives to the pro-rata order 
	matching algorithm used in stock exchanges. The main part of our work is to comprehensively consider various well known proportional representation methods and analyse 
	in details their suitability for replacing the pro-rata algorithm. Our analysis consists of a theoretical study as well as simulation studies based on data sampled from a 
	distribution which has been suggested in the literature as models of limit orders. Based on our analysis, we put forward the suggestion that the well known 
	Hamilton's method is a superior alternative to the pro-rata algorithm for order matching applications. 
\end{abstract}

\newpage

\section{Introduction\label{sec:intro}}
Financial instruments, such as shares of a company, are traded on stock exchanges. Traders place orders to buy and sell such instruments. 
An order specifies, among other things, the quantity or size, i.e.~the number of units to be purchased or sold. 
Orders can be of different types. 
A common and important type of order is a limit order which specifies both the quantity and the price at which the order is to be executed.
The quoted price is required to be a multiple of a unit of price called tick. A stock exchange maintains an order book which records for each financial instrument
the corresponding list of orders. The orders quoting the same buying or selling price are placed at the same price level in the order book. 

Trading in a stock exchange occurs by executing orders. Buy orders for an instrument are matched with corresponding sell orders. 
Matching happens in units of the instrument. The quantity specified in an order may not be equal to the quantity of the incoming counter-party order that it is matched with. An 
order is filled when all its units have been matched with one or more incoming counter-party matching orders. Unmatched or partially filled orders rest in the 
order book waiting for new orders to be matched with.

As a simple example, let us consider possible orders for the shares of a company. For simplicity, we assume that a limit order is specified by the order number (denoting the
chronology of appearance of the orders),
the type (${\sf b}$ for buy and ${\sf s}$ for sell), the size and the price. Suppose 
$(1,{\sf b},10,\$100)$,
$(2,{\sf b},50,\$100)$,
$(3,{\sf s},90,\$100)$,
$(4,{\sf b},20,\$110)$,
$(5,{\sf s},10,\$100)$,
$(6,{\sf s},40,\$100)$,
$(7,{\sf b},70,\$100)$ are seven such orders. The first two orders are buy orders and are stored in the order book as resting orders at price level $\$100$. The 
third order is a sell order at price $\$100$ whose size is 90, while the total size of the two resting buy orders in the order book at level $\$100$ is 60. So these two buy 
orders are satisfied by 60 units from the sell order. The leftover 30 units of the third order is stored as a resting order $(3,{\sf s},30,\$100)$ in the order book at price
level $\$100$. The fourth order is a buy order at price $\$110$ and is stored in the order book as a resting order at level $\$110$. The fifth and the sixth orders are sell 
orders at price $\$100$ and are also stored in the order book as resting orders at level $\$100$. The seventh order is a buy order at price $\$100$. At this point, the order 
book has three resting sell orders at level $\$100$, namely $(3,{\sf s},30,\$100)$, $(5,{\sf s},10,\$100)$, and $(6,{\sf s},40,\$100)$. The total size of the three 
resting sell orders is 80, while the incoming buy order has size 70. So all the resting orders cannot be satisfied. A method is required to determine how to
allocate the 70 units of the incoming buy order to the three resting orders. We come back to this example after discussing the more general setting.

%
Let us assume that at a particular price level there are $n$ resting orders, and the quantities of the orders are given by a vector $\mathbf{T} = (T_1, \ldots, T_n)$, 
where $T_i$ is a positive integer which represents the quantity of the $i$-th order. These $n$ orders are either all buy orders or all sell orders. 
Let $T = T_1 + \cdots + T_n$ be the total quantity of these orders. A new incoming counter-party order of quantity $S$ at the same price level
will be matched with the resting orders in $\mathbf{T}$. As a result, some or all of the resting orders may be executed. If $S \geq T$, the resting orders 
are all filled and can be fully executed. However, if $S < T$, not all resting orders can be completely filled. In this case, an order matching algorithm is required
to allocate portions of the incoming counter-party order to the $n$ resting orders. Henceforth, we will assume that the condition $S<T$ holds.

Formally, an order matching algorithm $\mathcal{M}(n, \mathbf{T}, S)$ takes as input the number $n$ of resting orders, the vector $\mathbf{T} = (T_1, \ldots, T_n)$ of the 
resting order quantities, and an incoming counter-party order of quantity $S$ with $0<S<T$. It 
outputs $\mathbf{S} = (S_1, \ldots, S_n)$ so that $S_i$ quantity of $T_i$ may be executed. In other words, $0\leq S_i \leq T_i$ and $S = S_1 + \ldots + S_n$. 

Two of the most common order matching algorithms are the price-time priority (also called first-come-first-served (FCFS) or
first-in-first-out (FIFO)) and the pro-rata methods (see for example~\cite{CMEMatchingAlgorithms,Pr11,CS2020,He22}). Both methods aim to achieve some kind of fariness
in the allocation of orders. In this work we focus on the pro-rata method.

The idea behind the pro-rata method is to distribute $S$ to the $n$ resting orders more or less in proportion to their fractions of the total order. 
So ideally the $i$-th resting order would receive $ST_i/T$ portion of the incoming counter-party order. This, however, has a problem. Financial instruments are
traded in indivisible atomic units. So if $ST_i/T$ is not an integer, then this amount of the order cannot be executed. 

The pro-rata order matching algorithm
adopts a two-step approach to this problem. In the first step, the algorithm assigns an amount $S_i^{\prime}=\lfloor ST_i/T\rfloor$ of the incoming counter-party order 
to the $i$-th resting 
order\footnote{For a real number $x$, $\lfloor x\rfloor$ denotes the greatest integer not greater 
than $x$, and $\lceil x\rceil$ denotes the least integer not less than $x$.}$^,$\footnote{Sometimes a simple modification to the first step is used. For 
example, the Chicago Mercantile Exchange (CME)~\cite{CMEMatchingAlgorithms} adopts the strategy in which if some $S_i^{\prime}$ turns out to be $1$, then it is instead set to 0.}.
This strategy consumes $S^{\prime}=S_1^{\prime}+\cdots+S_n^{\prime}$ units of the incoming counter-party order. In the second step, the remaining 
$S-S^{\prime}$ units are distributed to the resting orders based upon some strategy. The CME adopts the FIFO strategy for the second step~\cite{CMEMatchingAlgorithms}.
Accordingly, we will also assume the FIFO strategy in the second step for the purpose of this work. 

Going back to the previous example, it is required to allocate 70 units of the incoming buy order among the three resting orders of sizes 30, 10 and 40. Following the
pro-rata algorithm, in the first step, 26, 8 and 35 units are allocated to the three resting orders. This consumes 69 units. The remaining unit is allocated on the FIFO
basis to the first resting order. So the final allocation consists of 27, 8 and 35 units to the three resting orders. After the matching, 
the order book is updated to store $(3,{\sf s},3,\$100)$, $(5,{\sf s},2,\$100)$, and $(6,{\sf s},5,\$100)$ as resting orders at price level $\$100$.

In this paper, we raise two questions.
\begin{enumerate}
	\item How well does the pro-rata order matching algorithm achieve its goal of distributing the incoming order to the resting orders in proportion to their
		fractions of the total order?
	\item Are there other algorithms which perform better than the pro-rata algorithm in achieving proportionality?
\end{enumerate}
To answer the above questions, we need a measure to assess the performance of an order matching algorithm in achieving proportionality. For $n$ resting
orders, with $\mathbf{T}=(T_1,\ldots,T_n)$ the vector of quantities of the resting orders, and $S$ the size of the incoming order, the ideal allocation, or
the ideal proportional distribution is given by the vector 
\begin{eqnarray}\label{eqn-ideal}
\mathbf{I}=(ST_1/T,ST_2/T,\ldots,ST_n/T). 
\end{eqnarray}
Suppose $\mathcal{A}$ is an order matching algorithm which on input $n$, $\mathbf{T}$ and $S$
produces the allocation vector $\mathbf{S}_{\mathcal{A}}=(S_1,\ldots,S_n)$ as output, where $S_i\leq T_i$ for $i=1,\ldots,n$, and $S_1+\cdots+S_n=S$. 
The distance between the vectors $\mathbf{I}$ and $\mathbf{S}_{\mathcal{A}}$ is a measure of the performance of the algorithm $\mathcal{A}$ in achieving
proportionality. The closer $\mathbf{S}_{\mathcal{A}}$ is to $\mathbf{I}$, the better is the performance of $\mathcal{A}$ in achieving proportionality. 

We consider two standard measures of distance between two vectors, namely the $L_1$ and the $L_2$ metrics defined as follows.
\begin{eqnarray*}
L_1(\mathbf{S}_{\mathcal{A}},\mathbf{I})=\sum_{i=1}^n \mid S_i - ST_i/T \mid, & & L_2(\mathbf{S}_{\mathcal{A}},\mathbf{I})=\left(\sum_{i=1}^n (S_i - ST_i/T)^2\right)^{1/2}.
\end{eqnarray*}
Using these two metrics, we can quantifiably answer the first question posed above. The metrics also provide a method to address the second question.
For an order matching algorithm $\mathcal{A}$, let $\ell_{1,\mathcal{A}}$ and $\ell_{2,\mathcal{A}}$ denote
$L_1(\mathbf{S}_{\mathcal{A}},\mathbf{I})$ and $L_2(\mathbf{S}_{\mathcal{A}},\mathbf{I})$ respectively. For two order matching algorithms $\mathcal{A}_1$
and $\mathcal{A}_2$, we say that $\mathcal{A}_1$ is $L_1$-better (resp.~$L_2$-better) than $\mathcal{A}_2$ if 
$\ell_{1,\mathcal{A}_1}<\ell_{1,\mathcal{A}_2}$ (resp.~$\ell_{2,\mathcal{A}_1}<\ell_{2,\mathcal{A}_2}$). In other words, $\mathcal{A}_1$ is 
$L_1$-better (resp.~$L_2$-better) than $\mathcal{A}_2$, if its output is closer to the ideal allocation with respect to the $L_1$ (resp.~$L_2$) metric.
Using the above terminology, we can rephrase the second question as follows. 
\begin{quote}
Is there an order matching algorithm $\mathcal{A}$ which is better than the pro-rata order matching algorithm with respect to either or both of the $L_1$ and $L_2$ metrics?
\end{quote}

\subsection{Seat Distribution in Electoral Systems \label{subsec-elec} }
To answer the above question, we visit the literature on proportional representation in electoral systems which is far removed from stock exchanges and
more generally the financial world. 
Proportional representation is the most common kind of electoral system where the seats are not contested individually. Instead, the total number of seats is allocated 
to the contesting parties in proportion to the number of votes they have won in the election.  Let us consider an election contested by $n$ parties over $K$ seats that are 
distributed using a proportional representation method. Let $V_j$ denote the number of votes won by party $j \in \{1, \ldots, n\}$ in the election. The electoral output 
is denoted by the vector $\mathbf{V} = (V_1, \ldots, V_n)$, and the total number of votes cast in the election is $V = V_1 + \cdots + V_n$. Suppose the total number
of seats to be distributed among the parties is $K$. A proportional representation method determines the seat allocation vector 
$\mathbf{K}=(K_1,\ldots,K_n)$ where $K_i$ is the number of seats allocated to the $i$-th party, and $K_1+\cdots+K_n=K$. Typically, the total number of seats $K$ is much 
smaller than the total number of votes $V$, i.e.~$K<V$. Further, it is reasonable to assume that in practice the number of seats allocated to the 
$i$-th party is at most the number of votes received by the party, i.e.~$K_i\leq V_i$. 

Formally, a proportional representation method is an algorithm $\mathcal{A}(n, \mathbf{V}, K)$ which takes as input the number $n$ of parties, the vote count vector 
$\mathbf{V} = (V_1, \ldots, V_n)$, and the number $K$ of seats to be distributed, where $0<K<V$. It outputs the seat allocation vector $\mathbf{K} = (K_1, \ldots, K_n)$
such that $0\leq K_i\leq V_i$ and $K_1+\cdots+K_n=K$. 

There is a large literature on electoral systems in general and proportional representation methods in particular. We refer the reader to~\cite{BY01,Norris2004,Herron2018}
for elaborate discussions on these topics. A number of proportional representation methods have been proposed in the context of electoral systems. 
These can be divided into two types, the highest averages method and the largest remainder method. Detailed descriptions of these types of methods
are given in Section~\ref{sec:order-matching}.
The most well known of the highest averages method are the Jefferson/D'Hondt (JD) and the Webster/Sainte-Lagu\"{e} (WS) methods. Both of these methods continue to be of 
active interest. See for example~\cite{HM08,GF17,Me19,FSS20}.
Among the largest remainder method, the two well known methods are the Hare and the Droop methods. The version of the
largest remainder method as proposed by Hare was earlier described by Hamilton and is also called Hamilton's method.

A basic requirement that may be expected from a proportional representation method is that it satisfies the quota rule. 
The quota rule requires that for each party $i$, the number of seats $K_i$ allocated to the $i$-th party is equal to either $\lfloor (KV_i)/V\rfloor$ or $\lceil (KV_i)/V\rfloor$. 
Since $KV_i/V$ is the ideal allocation to the $i$-th party, the quota rule stipulates that for each party the deviation from the ideal is less than 1.
The pro-rata method as well as Hamilton's and Droop's methods satisfy the quota rule. 

Balinski and Young~\cite{BY01} identified certain paradoxes of proportional representation methods. A celebrated result by them states that if the number of
parties $n$ is at least 4 and the number of seats is at least $n+3$, then any proportional representation method which satisfies the quota rule must suffer
from the paradoxes. Since Hamilton's method satisfies the quota rule, it exhibits the paradoxes. Note that the pro-rata method also satisfies the quota rule
and hence must also exhibit the paradoxes, though this has not been reported earlier. The JD and the WS methods, on the other hand, are free of the paradoxes, and
so by the Balinski-Young theorem, must violate the quota rule. 

\subsection{Order Matching from Proportional Representation \label{subsec-om-pr}}

From the above description, it becomes clear that the goal of allocating an incoming counter-party order to resting orders in proportion to the sizes of the resting orders 
is the same as the task of assigning a fixed number of seats to several contesting parties in proportion to the number of votes obtained by these parties. The following
correspondence makes this observation precise.
\begin{center}
\begin{tabular}{lcl}
	$n$: number of resting orders & $\mapsto$ & $n$: number of parties, \\
	$S$: size of the counter-party order & $\mapsto$ & $K$: total number of seats, \\
	$T_i$: size of the $i$-th resting order & $\mapsto$ & $V_i$: number of votes polled by a party, \\
	$S_i$: allocation to the $i$-th resting order & $\mapsto$ & $K_i$: number of seats allocated to the $i$-th party.
\end{tabular}
\end{center}
With the above correspondence in mind, any algorithm for proportional representation of seats
in an electoral system becomes a potential candidate for use as an order matching algorithm by a stock exchange for proportional fulfillment of orders. The 
identification of proportional representation methods as possible substitutes for the pro-rata order matching algorithm is a key observation of the present work.
Not all proportional representation methods, however, are suitable for use as order matching algorithms. 
The main part of the present work is to make a comprehensive study of the important proportional 
representation methods available in the literature and determine their efficacies to the order matching problem.

The notions of ideal allocation and $L_1$ and $L_2$-distances to the ideal allocation were introduced in Section~\ref{sec:intro} in the context of order matching. Using the
above mentioned correspondence between order matching and proportional representation, these notions carry over directly to the context of proportional representation.
Hamilton's method has the attractive property that allocations produced by this method minimise the $L_1$ and $L_2$-distances to the ideal allocation, where
the minimum is over all possible allocation methods, i.e.~allocations produced by Hamilton's method enjoy optimality with respect to minimising the 
$L_1$ and $L_2$-distances to the ideal allocation.

As discussed above, there are several well known proportional representation 
methods which need to be compared in the context of order matching. There are several aspects to the comparison.
\begin{enumerate}
	\item What are the time complexities of the different proportional representation methods, and how do these compare to the pro-rata method?
	\item Some of the proportional representation methods have a positivity constraint, i.e.~they ensure that each party gets at least one seat.
		What is the relevance of such a positivity property in the context of order matching?
	\item What is the relevance of the paradoxes identified by Balinski and Young to the order matching scenario?
	\item How do the $L_1$ and $L_2$-distances between the ideal allocation and the allocations produced by the pro-rata and other proportional representation methods
		compare to the corresponding optimal distances (i.e.~the corresponding distances between the ideal allocation and the allocations produced by Hamilton's method)?
	\item How often and by how much do the JD and the WS methods violate the quota rule?
\end{enumerate}
We systematically answer the above questions. For the first question, we show that the pro-rata method is the fastest among the four methods, Hamilton's method
is only marginally slower than the pro-rata method, while the JD and the WS methods can require substantially more time. We argue that the second issue implies
a constraint which is hard to enforce in the context of order matching and more generally is not relevant in the financial world.
For the third issue, we argue that the paradoxes described by Balinski and Young are not relevant to the order matching scenario. This is because while 
in principle they do apply to the order matching scenario, in practice they are very unlikely to be actually encountered.
We tackle the fourth and the fifth questions by obtaining results
from simulation experiments. For the experiments, we followed the suggestion in~\cite{lim-bk2016} to model the generation of order book data.
From the results of the simulations, 
it turns out that the $L_1$-distances between the ideal allocation and the allocations produced by the pro-rata, the JD and the WS methods are respectively 
about 1.6, 2.3 and 1.1 times the $L_1$-distance between the ideal allocation and the allocations produced by Hamilton's method. Similar figures are obtained for the $L_2$-distance. 
Since the WS method is closer to the optimal (i.e.~Hamilton's method) than the JD method, 
we investigated quota rule violation for only the WS method. It has been mentioned in the influential work by Balinski and Young~\cite{BY01} that the WS method
violates the quota rule very rarely. Our experiments, however, show a different result. We find that quota rule violation by the WS method is quite frequent
and further the extent of violation is also non-negligible. A possible explanation for the difference of our experiments with what is reported in the literature
is that we have run the WS method on data which follows a distribution of resting orders that has been suggested in the order matching literature, while previous
work~\cite{BY01} on quota violation by the WS method considered data that arose in the context of electoral systems. 

In view of our analysis and experimental results, we find that Hamilton's method is a better alternative to the pro-rata method in the context of order matching. 
While both the pro-rata and Hamilton's method satisfy the quota rule, 
allocations produced by Hamilton's method minimise the $L_1$ and $L_2$-distances to the ideal allocation. So in terms of achieving best possible proportionality,
Hamilton's method achieves optimality among all possible order matching algorithms. In terms of efficiency, Hamilton's method is only marginally slower than the pro-rata method. 
In theory, Hamilton's method (and also the pro-rata method) suffers from certain paradoxex, but such paradoxes are extremely unlikely to occur in the practical 
context of order matching. 

\subsection{Procedural Fairness \label{subsec-fair} }
A recent paper by Hersch~\cite{He22} investigated the issue of procedural fairness of order allocation methods. In the paper, it was argued that
both the FIFO and the pro-rata are fair in principle, but not in practice. It was pointed out that the main disadvantage of pro-rata is the requirement
of the second step ``requiring exchanges to introduce secondary matching rules that can be gamed''. 

An alternative method called the random selection for service (RSS) method was proposed. 
Given the vector $\mathbf{T}=(T_1,\ldots,T_n)$ of resting orders, the RSS method defines a probability distribution $\pi$ over $\{1,\ldots,n\}$, where 
$\pi$ associates probability $T_i/T$ to $i$, for $i=1,\ldots,n$, 
Suppose the incoming counter-party order consists of $S$ units. Allocation is done by repeating the following procedure $S$ times:
an independent random $i$ is drawn from $\{1,\ldots,n\}$ following the probability distribution $\pi$ and
one unit is allocated to the $i$-th resting order. At the end of the procedure, let $S_i$ be the number of units allocated to the $i$-th resting order so that the
final allotment is $\mathbf{S}=(S_1,\ldots,S_n)$ satisfying $S=S_1+\cdots+S_n$.
It was argued by Hersch~\cite{He22} that the RSS method is fair in both principle and practice. Below we revisit this method and point out a crucial difference between principle
and practice.

Given $\mathbf{T}=(T_1,\ldots,T_n)$ and $S$, suppose the RSS method is executed $\Gamma$ times and for $\gamma=1,\ldots,\Gamma$, let
the allotment of $\gamma$-th execution be $\mathbf{S}_{\gamma}=(S_{\gamma,1},\ldots,S_{\gamma,n})$. For $i=1,\ldots,n$, let 
$\widehat{S}_i = (S_{1,i}+\cdots+S_{\Gamma,i})/\Gamma$, i.e.~$\widehat{S}_i$ is the average allotment to the $i$-th resting order computed over all the
$\Gamma$ trials. The law of large numbers assures us that asymptotically, i.e.~as $\Gamma$ goes to infinity, the average allotment $\widehat{S}_i$
tends to $ST_i/T$ which is equal to the $i$-th component of the ideal allocation vector $\mathbf{I}$ (see~\eqref{eqn-ideal}). So in principle, Hersch~\cite{He22} implicitly
considers achieving allocation close to the ideal allocation vector $\mathbf{I}$ to be procedurally fair. To this extent, Hersch's objective and ours coincide.
Additionally, our use of the $L_1$ and $L_2$ metrics to measure deviation from the ideal allocation vector can be considered to be a quantification of procedural fairness.
As such it expands the theoretical framework for studying procedural fairness of the order allocation methods.

From a practical point of view, however, the RSS method has a significant shortcoming. The law of large numbers applies in an asymptotic context, i.e.~as
$\Gamma$ goes to infinity. In practice, given $\mathbf{T}=(T_1,\ldots,T_n)$ and $S$, a stock exchange will execute the RSS method exactly once to obtain a single allocation 
vector $\mathbf{S}=(S_1,\ldots,S_n)$. In other words, in practice the value of $\Gamma$ will be 1. 
The law of large numbers does not say anything about the value obtained in a single execution. 
As a result, the allocation obtained from a single execution may not turn out to be fair. 
The $S_i$'s obtained after a single execution of RSS can be any value in the set $\{0,\ldots,S\}$. In particular, there is a positive probability that for
some $i$, $S_i$ turns out to be greater than $T_i$. 
Considering a particular example with $n=2$, $\mathbf{T}=(10,90)$ and
$S=10$, Hersch~\cite{He22} provides probabilities that the $S_i$'s can take certain values: the probability that $S_1\geq 1$ (resp.~$S_1=10$) is about 
0.88 (resp.~$7\times 10^{-6}$); the probability that $S_2=20$ (resp.~$S_2\geq 10$) is about 0.12 (resp.~approaches 1). These probabilities, however, do not
enlighten us about the concrete values of the $S_i$'s after a single execution. In particular, the probability that $S_2\geq 10$ approaches 1 suggests an
asymptotic context, where the frequentist view of probability is taken. To interpret such probabilities, one again needs to consider a large number $\Gamma$
of trials of the RSS method and consider the average allocation over all the $\Gamma$ trials. 

To test the practical efficacy of the RSS method, we have run experiments with the method. It turns out that the allocation vector obtained by the RSS method
has a very large deviation from the ideal allocation vector in terms of both the $L_1$ and the $L_2$ metrics. Particular examples are provided in Section~\ref{sec-sim-res}.
By the above explanation, this observation is not surprising.

As mentioned earlier, according to Hersch, the main disadvantage of the pro-rata method is the use of secondary matching rules in the second step of the method
which leads to the possibility of gaming. The second step of Hamilton's method is completely determined and cannot be gamed. 
So our proposal of using Hamilton's method overcomes the disadvantage of the pro-rata method pointed out by Hersch. In terms of procedural fairness as measured by
distance to the ideal allocation vector, Hamilton's method achieves the optimal distance.

\subsection{Contribution to Proportional Representation \label{subsec-prop-rep-cont}}
Apart from the investigation of the applicability of proportional representation methods to the order matching problem, our work provides new insights into 
methods for proportional representation which are of independent interest. 

The first such insight is to suggest the use of priority queue data structures to efficiently implement
various proportional representation algorithms and analyse their time complexities. To the best of our knowledge, such an exercise has not been done earlier in the
literature and would be of interest to various applications of proportional representation methods. 

The second insight is that the WS method violates quota
quite often for the data generated from the distributions which model the manner in which orders appear in a stock exchange. This contradicts the prior reported
observation that the WS method rarely violates quota. As mentioned earlier, the previous observation is based on data generated from distributions which model voting patterns. 
By showing an example of a distribution for which the WS method violates quota often, we open up the possibility of investigating whether there are other distributions for which
the WS method also frequently violates quota. Understanding and identifying such distributions will be important for suggesting the use of the WS method for applications
beyond determining seats in electoral systems.

\section{Proportional Representation Methods \label{sec:order-matching}}
There are several different proportional representation methods (see~\cite{BY01,Norris2004,Herron2018}). Such methods are also called apportionment methods.
Below we describe two well known families of proportional representation methods, namely the highest averages or the divisor method, and the highest remainder method. 

Recall the setting described in Section~\ref{subsec-elec}, where $V_i$ is the number of votes received by the $i$-th party and $K$ is the total number of available seats. 
The goal is to determine $K_i$ which is the number of seats allocated to the $i$-th party.

\paragraph{Highest averages method.}
Let $f: \mathbb{Z}^{+} \cup \{0\} \to \mathbb{R}$ be a function from the non-negative integers to the reals. Let $V=V_1+\cdots+V_n$ and $v_i=V_i/V$. 
In the highest averages method (also called divisor method), seats are allocated iteratively. The seat distribution algorithm goes through $K$ iterations and 
in each iteration exactly one seat 
is allocated to one of the parties. Initially, the algorithm sets $K_1=K_2=\cdots=K_n=0$. For $k$ from $1$ to $K$, in the $k$-th iteration
the algorithm determines $j=\arg\max\{V_i/f(K_i): i=1,\ldots,n\}$ and increments $K_j$ by one. After $K$ iterations, the final values of $K_1,\ldots,K_n$ are the numbers of
seats allocated to the various parties. Various different methods arise from the different definitions of $f$. The definitions of $f$ for several well known
methods including the popular Jefferson/D'Hondt and the Webster/Sainte-Lagu\"{e} methods are shown in Table~\ref{tab:highest-averages-methods}.

\begin{table}[!htb]
    \centering
    \begin{tabular}{l|l}
    \hline
        Method name & $f(t)$ \\
    \hline
	    Jefferson/D'Hondt (JD) & $(t+1)$ \\
	    Webster/Sainte-Lagu\"{e} (WS) & $(t+0.5)$ \\
	    Adam (Ad) & $\lceil t \rceil$ \\
	    Danish (Da) & $t/(t+(1/3))$ \\
	    Dean (De) & ${t(t+1)}/{(t+0.5)}$ \\
	    Huntington/Hill (HH) & $\sqrt{t(t+1)}$ \\
    \hline
    \end{tabular}
	\caption{Expressions for the function $f$ corresponding to some well known highest averages method. \label{tab:highest-averages-methods}}
\end{table}

\paragraph{Largest remainder method.}
The method uses a parameter $Q$. The seat allocation is done in two phases. In the first phase, the $i$-th party is allocated $\lfloor V_i/Q\rfloor$ 
seats. Let $R_i=V_i/Q - \lfloor V_i/Q\rfloor$ be the remainder corresponding to the $i$-th party. Suppose after the first phase $k$ seats remain unallocated. 
In the second phase, the parties with the $k$ largest remainders are each allocated one seat. Let $K_i$ be the number of seats allocated to the $i$-th party at the
end of the second round. The method ensures that $K_1+\cdots+K_n=K$. 
Table~\ref{tab-LR} shows the expressions for $Q$ corresponding to two well known methods.

\begin{table}[!htb]
    \centering
    \begin{tabular}{l|l}
    \hline
        Method name & $Q$ \\
    \hline
	    Hare (Ha) & $V/K$ \\
	    Droop (Dr) & $1 + \lfloor V/(1+K) \rfloor$ \\
    \hline
    \end{tabular}
	\caption{Expressions for the parameter $Q$ corresponding to two well known largest remainder method. \label{tab-LR}}
\end{table}
The largest remainder method with Hare's value for $Q$ was originally described by Hamilton and is also called Hamilton's method. The abbreviation Ha stands
for both Hamilton and Hare.

The largest remainder method satisfies the quota rule, i.e.~the condition that $K_i$ is equal to either $\lfloor K V_i/V\rfloor$ or $\lceil K V_i/V\rceil$ is achieved
by allocations produced by the largest remainder method.
It is known that the allocations produced by Hamilton's method minimise the $L_1$ and $L_2$-distances to the ideal proportional seat allocation $(KV_1/V,KV_2/V,\ldots,KV_n/V)$,
where the minimum is over all possible apportionment methods.
In fact, the allocations produced by Hamilton's method minimise $L_p$-distance to the ideal allocation for any finite $p\geq 1$. See~\cite{Bi76}
and also Proposition~3.7 on Page~117 of~\cite{BY01}. 

\paragraph{Paradoxes.}
Balinski and Young~\cite{BY01} identified certain paradoxes of apportionment methods. A scenario (or setting) of the proportional representation problem
is described by the vector of obtained votes and the total number of seats. Consider two possible scenarios, where $(V_1,\ldots,V_n)$ and $K$ represents the
first scenario and $(V_1^{\prime},\ldots,V_n^{\prime})$ and $K^{\prime}$ represents the second.
Let $(K_1,\ldots,K_n)$ and $(K_1^{\prime},\ldots,K_n^{\prime})$ be the seat allocation vectors corresponding to the two scenarios. 
\begin{itemize}
	\item The Alabama paradox is the following: $V_i=V_i^{\prime}$ for $i=1,\ldots,n$
		and $K^{\prime}>K$, but there is a $j\in \{1,\ldots,n\}$ such that $K_j^\prime<K_j$. 
		In other words, the votes polled remain the same and the total number of seats has gone up, but the number of seats allocated to a party has gone down.
	\item The population paradox is the following: $K=K^{\prime}$ and for distinct $i,j\in\{1,\ldots,n\}$, $V_i^\prime/V_j^\prime > V_i/V_j$, but
		$K_i^{\prime}<K_i$ and $K_j^\prime>K_j$.
		In other words, the number of seats remains the same while the ratio of the number of votes polled by the $i$-th party to the number of votes 
		polled by the $j$-th party increases, yet the seat allocation to the $i$-th party decreases, and the seat allocation to the $j$-th party
		increases.
\end{itemize}
Balinski and Young~\cite{BY01} defined a property, called {\em population monotone}, which states that if the $i$-th party's vote increases relative to the
$j$-th party, then the $i$-th party should not get fewer seats and the $j$-th party more. They showed that if any apportionment method satisfies this property, then it does
not exhibit either the Alabama or the population paradox. They further showed that an apportionment method is population monotone if and only if it is a divisor method. 
A famous impossibility result due to Balinski and Young is that for $n\geq 4$ and $K\geq n+3$, there is no apportionment method which is both population monotone and 
satisfies the quota rule. 

Since the JD and the WS methods are divisor methods, it follows from the above mentioned results that they are population monotone. As a consequence, on the positive side, 
these two methods do not exhibit either the Alabama or the population paradoxes, while on the negative side they do not satisfy the quota rule. Balinski and Young's results also 
show that since the pro-rata and Hamilton's methods satisfy the quota rule, they are not population monotone and consequently must exhibit the paradoxes. There are known
examples of the paradoxes for Hamilton's method and it is easy to get similar examples for the pro-rata method. 
The largest remainder method with the Droop's choice for $Q$ satisfies the quota rule and consequently exhibits the above paradoxes. Further, unlike Hamilton's
method, allocations produced by Droop's method do not achieve optimal distance to the ideal allocation. So in the context of order matching, there is no reason
to consider Droop's method in comparison to Hamilton's method.

\section{Alternatives to Pro-Rata \label{sec-alt}}
We consider in detail the various aspects of the comparison between the various methods which were identified in Section~\ref{subsec-om-pr}. Three of these are theoretical
and are discussed in this section, while the other two are based on experimental results and are discussed in the next section.

In this section and the next, let $n$ be the number of resting orders, $\mathbf{T}=(T_1,\ldots,T_n)$ be the vector of resting orders and $S$ be the size of the
incoming counter-party order. The allocation vector is $\mathbf{S}=(S_1,\ldots,S_n)$. 

\subsection{Computational Efficiency \label{subsec-comp}}
The first step of the pro-rata method computes $S_i^\prime=\lfloor ST_i/T\rfloor$ for $i=1,\ldots,n$. So this step requires $O(n)$ time. Suppose that the vector
$$\mathbf{S}^{\prime}=(S_1^{\prime},S_2^{\prime},\ldots,S_n^{\prime})=(\lfloor ST_1/T \rfloor, \lfloor ST_2/T \rfloor, \ldots, \lfloor  ST_n/T\rfloor)$$
is obtained after the first step and let $S^{\prime}=S_1^{\prime}+S_2^{\prime}+\cdots+S_n^{\prime}$. Let $r=S-S^{\prime}$. Then in the second step $r$ units need to be
allocated using FIFO. Note that $r\leq n$. So the FIFO strategy of the second step requires $O(r)=O(n)$ time. Consequently, the overall
complexity of the pro-rata method is $O(n+r)=O(n)$ time.

The computation of the JD and the WS methods (and more generally of the highest averages methods) differ only in the choice of the function $f$. This choice does not 
affect the complexity of the algorithm. The algorithm 
performs $S$ iterations. In each iteration it needs to find a maximum among $n$ numbers. So if the algorithm is implemented in a straighforward manner, then the time complexity 
is $O(Sn)$. This can be improved to $O(n+S\log n)$ by using an appropriate data structure as we discuss next.

A priority queue or a max-heap~\cite{AHU78} is a data structure satisfying the following properties. 
For $n$ numbers, a max-heap can be built in $O(n)$ time such that the maximum of these numbers is at the top of the heap. 
Further, if any one of the $n$ numbers is changed, then using only $O(\log n)$ operations, the new maximum can be placed on the top of the heap. 

The JD and the WS methods start with an initial allocation
$\mathbf{U}=(U_1,\ldots,U_n)$, with $U_i=0$ for $i=1,\ldots,n$. In each of the $S$ iterations, the vector $\mathbf{U}$ is updated by incrementing exactly one
component. The $j$-th component of $\mathbf{U}$ is updated where $j=\arg\max\{T_i/f(U_i): i=1,\ldots,n\}$ (ties can be broken arbitrarily, or in a FIFO manner). 
At the end of $S$ iterations, the value of $\mathbf{U}$ provides the final allocation vector $\mathbf{S}=(S_1,\ldots,S_n)$. 
The implementation of these methods can benefit from the use of a max-heap. The pairs $(i,T_i/f(U_i))$, $i=1,\ldots,n$ are stored such that
the numbers $T_i/f(U_i)$, $i=1,\ldots,n$ constitute a max-heap. Initially building this max-heap requires $O(n)$ time. In any iteration, the top of the heap 
provides the pair $(j,T_j/f(U_j))$ such that $j=\arg\max\{T_i/f(U_i): i=1,\ldots,n\}$. The $j$-th component of $\mathbf{U}$ is incremented. As a result, 
of the $n$ numbers $(i,T_i/f(U_i))$, $i=1,\ldots,n$ only the $j$-th number changes, while all the other numbers remain unchanged. Next, heap operations are done
to ensure that maximum of the updated numbers is on the top of the heap. This requires $O(\log n)$ time. So building the heap requires $O(n)$ time while
each of the $S$ iterations requires $O(\log n)$ time. So the total time required is $O(n+S\log n)$.

In Adam's, Dean's and Huntington/Hill's methods as well as the Danish method, it is required to start by allocating one unit to each of the $n$ 
resting orders (see Section~\ref{subsec-positivity} for an explanation). So a total of $S-n$ iterations are required to allocate the remaining $S-n$ units of 
the incoming counter-party order. Consequently, the time complexity of these four methods is $O(n+(S-n)\log n)$. 

For Hamilton's method (and more generally for the largest remainder method), the first step is the same as the first step of the pro-rata method and consequently 
requires $O(n)$ time. In the second
step, the leftover $r=S-S^{\prime}\leq n$ units are to be allocated to the resting orders having the $r$ largest remainders. This requires determining the
$r$ largest remainders. A simple way to perform the second step is to compute all the $n$ remainders, i.e.~$R_i=ST_i/T - \lfloor ST_i/T \rfloor$, $i=1,\ldots,n$ and 
then sort the $R_i$'s in descending order. Next assign one unit to each of the resting orders corresponding to the first $r$ entries in the sorted list of
remainders. Sorting requires $O(n\log n)$ time and so the time complexity of Hamilton's method is dominated by the
sorting step and the overall time complexity of the method is $O(n+n\log n)=O(n\log n)$. 

The idea of using a max-heap can be used to improve the efficiency of Hamilton's method. After the first step, the pairs $(i,R_i)$, $i=1,\ldots,n$ are
stored such that the numbers $R_1,\ldots,R_n$ form a max heap. This requires $O(n)$ time. Suppose $(j,R_j)$ is the entry on the top of the heap. 
Then $R_j$ is the largest remainder. So $S_j^\prime$ is incremented and the remainder $R_j$ is updated to the value obtained by subtracting the new value of $S_j^\prime$
from $ST_j/T$. All other remainders remain unchanged. Next, heap operations are done to ensure that the maximum gets on the top. This requires $O(\log n)$ time. 
Since $r$ leftover units have to be allocated, and the allocation of each of these $r$ units requires $O(\log n)$ time, the second step requires 
a total of $O(r\log n)$ time. So using a max-heap, the two steps of Hamilton's method can be implemented in $O(n+r\log n)$ time. 

Table~\ref{tab-eff-comp} compares the time complexities of the above methods. Clearly pro-rata is the fastest. Since $r$ is at most $n$, Hamilton's method
(and also Droop's method) requires at most $O(n\log n)$ time. This is more than the time required by the pro-rata method by a factor of $\log n$. So Hamilton's method is slower than
pro-rata by only a logarithmic factor. The JD and the WS methods require time $O(n+S\log n)$. Here $S$ is the size of the incoming counter-party order. Note that the
size of the incoming counter-party order has no relation to the number of resting orders, i.e.~$n$. For typical order matching applications, $S$ may be several orders
of magnitude larger than $n$. In such situations, the JD and the WS methods will require much more time than the pro-rata method. 
So from the viewpoint of efficiency, Hamilton's method will be marginally slower than pro-rata, while both the JD and the WS methods may turn out to be
significantly slower. In electronic stock exchanges where orders are executed extremely fast, the slowdown due to the JD and the WS methods may not be acceptable.

\begin{table}
\centering
	\begin{tabular}{|c|c|c|c|}
		\hline 
		pro-rata & Ha or Dr & JD or WS & Ad, Da, De, or HH \\ \hline
		$O(n+r)$ & $O(n+r\log n)$ & $O(n+S\log n)$ & $O(n+(S-n)\log n)$ \\ \hline
	\end{tabular}
	\caption{Time complexities of the different methods. In the table, $r\leq n$ is the number of leftover units after the first step of the pro-rata
	algorithm. \label{tab-eff-comp}}
\end{table}

\subsection{Relevance of Positivity Requirement \label{subsec-positivity}}
Apart from the JD and the WS methods, Table~\ref{tab:highest-averages-methods} lists several other methods which fall within the class of the highest averages method, namely
Adam, Danish, Dean and Huntington/Hill.
Note that $f(t)$ for these methods satisfy the condition $f(0)=0$. 
Suppose we start the procedure with initial seat allocations to be 0, i.e. $K_i=0$ for $i=1,\ldots,n$. Then in the first iteration, each $f(K_i)$ is 0 and
so the computation for determining $j=\arg\max\{V_i/f(K_i): i=1,\ldots,n\}$ will encounter divisions by zero and the method will fail. 
This problem is bypassed by starting the procedure with an initial allocation of one seat to each party. So at the
end of the procedure, each party is ensured to have at least one seat. 
Note that the constraint of allocating at least one seat to each party requires the number of seats to be at least as large as the
number of parties. 

In the context of order matching, the feature of assigning at least one seat to each party translates to assiging at least one unit of the 
incoming counter-party order to each of the resting orders. This requires the size of the incoming counter-party order to be at least the number of resting orders,
i.e.~$S\geq n$. While this condition is likely to hold, it cannot be imposed in general, since there is no control over the size $S$ of the incoming counter-party order. 
More generally, the principle of assigning at least one unit of the incoming order to each of the resting orders does not have any justification in the context
of stock exchanges. On the contrary,
some stock exchanges follow the rule that if the order unit determined by the first step of the pro-rata method is 1, then this is instead set to 0~\cite{CMEMatchingAlgorithms}.
The principle of at least one unit for each resting order may not be welcome by such exchanges. Due to this reason as well as the unimplementable constraint
of $S\geq n$, the methods due to Adam, Danish, Dean and Huntington/Hill are not appropriate for the order matching application.

\subsection{Relevance of the Paradoxes \label{subsec-paradox}}

Let us consider the relevance of the the Alabama and the population paradoxes in the context of order matching. 
The paradoxes refer to two scenarios involving the sizes of the resting orders and the size of the incoming counter-party order.

The Alabama paradox translates to the following. The sizes of the resting orders remain the same in both the first and the second scenarios, while
an increase occurs in the incoming counter-party order of the second scenario in comparison to the first scenario. Yet the allocation to a particular resting order goes down
in the second scenario compared to the first. 

The population paradox translates to the following. The size of the incoming counter-party order remains the same in both the scenarios, and there are two
resting orders $i$ and $j$ such that the size of the $i$-th order relative to the size of the $j$-th order increases in the second scenario compared to the first. 
Yet the allocation to the $i$-th order decreases while the allocation to the $j$-th order increases.

From the point of view of practical order matching scenarios, the above paradoxes are unlikely to arise. For the paradoxes to occur, some quantity must remain
the same in the two scenarios. For the Alabama paradox, the sizes of the resting orders remain the same in the two scenarios, while for the population paradox, the size of the
incoming counter-party order is the same in the two scenarios. Since both the resting orders and the incoming counter-party order are generated by random processes, it is 
very unlikely that in any two real-life scenarios, either of these will remain the same. So even though the paradoxes apply in theory to the pro-rata and the Hamilton methods, 
they are very unlikely to be encountered in practice.

\section{Simulation and Empirical Results \label{sec-sim-res}}

In Table~\ref{tab-ex}, we provide some examples of the order allocation vector $\mathbf{S}_{\mathcal{A}}$, where 
$\mathcal{A}$ is one of $\mathcal{P}$ (denoting the pro-rata method), RSS, Hamilton (Ha), JD, or WS methods. 
The ideal allocation vector is $\mathbf{I}=(ST_1/T,\ldots,ST_n/T)$. 
From the examples, we observe that for the RSS method, the $L_1$ and $L_2$ distances from the ideal allocation vector $\mathbf{I}$ are much larger than these
distances from the other methods. This is as expected (see Section~\ref{subsec-fair}) and highlights the impracticability of the RSS method. While the table
provides only three examples, we have obtained many other examples where the observation that the order allocation vector produced by the RSS method
is much farther away from the ideal allocation vector compared to the other methods holds. In view of this, we do not consider the
RSS method any further in our simulation studies.

\begin{table}
\centering
{\scriptsize
	\begin{tabular}{|l|l|r|r|r|r|r|r|r|r|r|r|r|r|}
		\cline{13-14}
		\multicolumn{12}{c|}{} & $\ell_{1,\mathcal{A}}$ & $\ell_{2,\mathcal{A}}$ \\ \hline
		\multirow{5}{*}{Ex~1}
		& $\mathbf{T}$ & 209 & 727 & 746 & 808 & 995 & 204 & 598 & 773 & 979 & 899 & - & - \\ \cline{2-14}
		& $\mathbf{I}$ & 3.01 & 10.48 & 10.75 & 11.65 & 14.34 & 2.94 & 8.62 & 11.14 & 14.11 & 12.96 & - & - \\ \cline{2-14}
		& $\mathbf{S}_{\mathcal{P}}$ & 4 & 11 & 11 & 12 & 15 & 2 & 8 & 11 & 14 & 12 & 5.54 & 2.02 \\ \cline{2-14}
		& $\mathbf{S}_{{\rm RSS}}$ & 4 & 7 & 17 & 11 & 18 & 3 & 5 & 10 & 15 & 10 & 23.69 & 9.48 \\ \cline{2-14}
		& $\mathbf{S}_{{\rm Ha}}$ & 3 & 10 & 11 & 12 & 14 & 3 & 9 & 11 & 14 & 13 & 2.17 & 0.85 \\ \cline{2-14}
		& $\mathbf{S}_{{\rm JD}}$ & 3 & 10 & 11 & 12 & 14 & 3 & 9 & 11 & 14 & 13 & 2.17 & 0.85 \\ \cline{2-14}
		& $\mathbf{S}_{{\rm WS}}$ & 3 & 10 & 11 & 12 & 14 & 3 & 9 & 11 & 14 & 13 & 2.17 & 0.85 \\ \hline
		\multirow{5}{*}{Ex~2}
		& $\mathbf{T}$ & 1 & 655 & 307 & 138 & 647 & 48 & 625 & 382 & 95 & 424 & - & - \\ \cline{2-14}
		& $\mathbf{I}$ & 0.03 & 19.72 & 9.24 & 4.15 & 19.48 & 1.44 & 18.81 & 11.50 & 2.86 & 12.76 & - & - \\ \cline{2-14}
		& $\mathbf{S}_{\mathcal{P}}$ & 0 & 20 & 10 & 5 & 20 & 0 & 18 & 11 & 2 & 12 & 6.82 & 2.44 \\ \cline{2-14}
		& $\mathbf{S}_{{\rm RSS}}$ & 0 & 21 & 15 & 2 & 21 & 1 & 17 & 8 & 4 & 11 & 19.41 & 7.87 \\ \cline{2-14}
		& $\mathbf{S}_{{\rm Ha}}$ & 0 & 20 &  9 &  4 & 19 & 1 & 19 & 12 &  3 & 13 & 2.69 & 0.97 \\ \cline{2-14}
		& $\mathbf{S}_{{\rm JD}}$ & 0 & 20 & 9 & 4 & 20 & 1 & 19 & 12 & 2 & 13 & 3.46 & 1.31 \\ \cline{2-14}
		& $\mathbf{S}_{{\rm WS}}$ & 0 & 20 & 9 & 4 & 19 & 1 & 19 & 12 & 3 & 13 & 2.69 & 0.97 \\ \hline
		\multirow{5}{*}{Ex~3}
		& $\mathbf{T}$ & 268 & 806 & 409 & 420 & 869 & 659 & 189 & 317 & 286 & 721 & - & - \\ \cline{2-14}
		& $\mathbf{I}$ & 5.42 & 16.30 & 8.27 & 8.50 & 17.58 & 13.33 & 3.82 & 6.41 & 5.78 & 14.58 & - & - \\ \cline{2-14}
		& $\mathbf{S}_{\mathcal{P}}$ & 6 & 17 & 9 & 9 & 18 & 13 & 3 & 6 & 5 & 14 & 5.86 & 1.92 \\ \cline{2-14}
		& $\mathbf{S}_{{\rm RSS}}$ & 2 & 15 & 8 & 4 & 12 & 17 & 2 & 6 & 9 & 25 & 34.61 & 14.16 \\ \cline{2-14}
		& $\mathbf{S}_{{\rm Ha}}$ & 5 & 16 &  8 &  9 & 18 & 13 & 4 & 6 &   6 & 15 & 3.47 & 1.14 \\ \cline{2-14}
		& $\mathbf{S}_{{\rm JD}}$ & 5 & 17 & 8 & 8 & 18 & 13 & 4 & 6 & 6 & 15 & 3.86 & 1.30 \\ \cline{2-14}
		& $\mathbf{S}_{{\rm WS}}$ & 5 & 16 & 8 & 9 & 18 & 13 & 4 & 6 & 6 & 15 & 3.47 & 1.14 \\ \hline
	\end{tabular}
	\caption{Examples of simulation runs with $n=10$ and $S=100$. \label{tab-ex}}
}
\end{table}

We have run more extensive simulation results to compare the four methods, namely pro-rata, Hamilton, JD and WS. It would have been more appropriate to run the simulations
on actual order book data obtained from some stock market. Unfortunately, we do not have access to such data. So instead we generated order book data
following recommendations available in the literature. Section~2.3 of the book~\cite{lim-bk2016} entitled ``Limit Order Book'' mentions that empirical 
studies show that the distribution of order sizes is complex to characterize and that a power-law distribution has been suggeted in the literature. 
For limit orders, the decay rate was suggested to be equal to approximately $2.0$. Further, orders tend to have a ``round'' size with clusters
around 100 and 1000 being observed. Based on this description, we used the power law distribution with decay rate 2.0 to generate the orders. A sample
obtained from the power law is a real number not less than 1. We rounded this sample to the nearest integer and multiplied this integer by the quantum size
$q$. This provided a particular resting order. To obtain a vector of $n$ resting orders, we independently repeated the procedure $n$ times. 
We used two values of the quantum size, namely $q=100$ and $q=1000$ to incorporate the ``rounding'' effect mentioned in~\cite{lim-bk2016}. For each value of 
$n$ and $q$, we obtained $N$ vectors of resting orders. Suppose $\mathbf{T}=(T_1,\ldots,T_n)$ is a vector of resting orders and let $T=T_1+\cdots+T_n$.
To apply the order matching algorithms, we need the value $S$ of the size of the incoming counter-party order. We wish to have $S<T$ so that trivial allocation does not
hold. The value of $S$ was chosen to be a random integer in the interval $[0,T-1]$. In our experiments, we have taken $N=1000$. The simulation parameters are given 
in Table~\ref{tab-sim-param}.

\subsection{Distance to the Ideal Allocation \label{subsec-d2ideal}}
Given $\mathbf{T}$ and $S$, let as before $\ell_{1,\mathcal{A}}$ and $\ell_{2,\mathcal{A}}$ be the $L_1$ and $L_2$-distances respectively
of the output of algorithm $\mathcal{A}$ to the ideal allocation, where $\mathcal{A}$ is one of pro-rata, Hamilton, JD or WS methods. From the discussion in
Section~\ref{sec:order-matching}, we know that $\ell_{1,{\rm Ha}}$ is the minimum $L_1$-distance to the ideal allocation and $\ell_{2,{\rm Ha}}$ is the minimum $L_2$-distance 
to the ideal allocation, where the minimum is taken over all algorithms $\mathcal{A}$. So for the other algorithms, it makes sense to compare the $L_1$-distance
to $\ell_{1,{\rm Ha}}$ and the $L_2$-distance to $\ell_{2,{\rm Ha}}$. For $\mathcal{A}$ in $\{\mathcal{P},{\rm JD},{\rm WS}\}$, and for 
$\ell_{1,\mathcal{A}}\neq 0 \neq \ell_{2,\mathcal{A}}$ we define
\begin{eqnarray}\label{eqn-rho}
	\rho_{1,\mathcal{A}}=\frac{\ell_{1,\mathcal{A}}}{\ell_{1,{\rm Ha}}} & \mbox{ and } & \rho_{2,\mathcal{A}}=\frac{\ell_{2,\mathcal{A}}}{\ell_{2,{\rm Ha}}}.
\end{eqnarray}
To measure the efficacy of algorithm $\mathcal{A}$ with respect to Hamilton's method, we decided to study the distributions of 
$\rho_{1,\mathcal{A}}$ and $\rho_{2,\mathcal{A}}$ using simulations. 

In each simulation, there were $N$ iterations and in each of these iterations a vector $\mathbf{T}=(T_1,\ldots,T_n)$ of resting orders and a incoming counter-party order $S$
were obtained as described above. For each pair $\mathbf{T}$ and $S$, we computed $\ell_{1,{\rm Ha}}$ and $\ell_{2,{\rm Ha}}$ and then 
$\rho_{1,\mathcal{A}}$ and $\rho_{2,\mathcal{A}}$ for $\mathcal{A}$ in $\{\mathcal{P},{\rm JD},{\rm WS}\}$.
At the end of $N$ iterations of the simulation, we computed the
mean $\mu_{1,\mathcal{A}}$ and standard deviation $\sigma_{1,\mathcal{A}}$ of the obtained values of $\rho_{1,\mathcal{A}}$ and similarly,
the mean $\mu_{2,\mathcal{A}}$ and standard deviation $\sigma_{2,\mathcal{A}}$ of the obtained values of $\rho_{2,\mathcal{A}}$.
The obtained values of the means and the standard deviations are shown in Table~\ref{tab-res}.
Each entry in the table is of the form $(\mu,\sigma)$ which provides the mean and the standard deviation of $\rho$ corresponding to the case identified by the
row and column labels. We have the following observations from the results. 
\begin{enumerate}
	\item For each $\mathcal{A}$ in $\{\mathcal{P},{\rm JD},{\rm WS}\}$, the values of $\mu_{1,\mathcal{A}}$ and $\mu_{2,\mathcal{A}}$ are almost the same, 
		as are the values of $\sigma_{1,\mathcal{A}}$ and $\sigma_{2,\mathcal{A}}$. Further, across simulations the means and the standard deviations
		are also very close. This suggests the stability of the results.
	\item The distance to the ideal for pro-rata is about 1.6 times the distance to the ideal for Hamilton's method. For the JD and the WS methods, these 
		numbers are about 2.2 and 1.1 respectively. We may conclude that among pro-rata, JD and the WS methods, the WS method produces allocations 
		which are closest to the optimal (i.e.~allocations produced by Hamilton's method), while the JD method produces allocations which are
		farthest from the optimal. Importantly, the allocations produced by the pro-rata method are significantly far from the optimal allocations.
\end{enumerate}
In view of the above observations, between the JD and the WS methods, we find no reason to consider the JD method. 

\begin{table}
\begin{subtable}{0.5\textwidth}
	\centering
	{\scriptsize
		\begin{tabular}{|l|r|r|}
			\cline{2-3}
			\multicolumn{1}{c|}{} & \multicolumn{1}{c|}{$n$} & \multicolumn{1}{c|}{$q$} \\ \hline
			Sim~1 & 50 & 100 \\ \hline 
			Sim~2 & 50 & 1000 \\ \hline 
			Sim~3 & 100 & 100 \\ \hline 
			Sim~4 & 100 & 1000 \\ \hline 
			Sim~5 & 150 & 1000 \\ \hline 
			Sim~6 & 200 & 1000 \\ \hline 
		\end{tabular}
		}
		\caption{Parameters of the various simulation runs. \label{tab-sim-param} }
\end{subtable}
\begin{subtable}{0.5\textwidth}
	\centering
	{\scriptsize
		\begin{tabular}{|l|l|r|r|r|}
			\cline{3-5} 
			\multicolumn{2}{c|}{} & \multicolumn{1}{c|}{pro-rata} & \multicolumn{1}{c|}{JD} & \multicolumn{1}{c|}{WS} \\ \hline
			\multirow{2}{*}{Sim~1} & 
			  $L_1$ & (1.63,0.19) & (2.23,1.08) & (1.12,0.12) \\ \cline{2-5}
			& $L_2$ & (1.64,0.18) & (2.23,1.08) & (1.12,0.12) \\ \hline
			\multirow{2}{*}{Sim~2} & 
			  $L_1$ & (1.62,0.19) & (2.22,0.94) & (1.12,0.12) \\ \cline{2-5}
			& $L_2$ & (1.63,0.19) & (2.22,0.94) & (1.12,0.12) \\ \hline
			\multirow{2}{*}{Sim~3} & 
			  $L_1$ & (1.64,0.14) & (2.41,1.19) & (1.15,0.13) \\ \cline{2-5}
			& $L_2$ & (1.65,0.14) & (2.41,1.19) & (1.15,0.13) \\ \hline
			\multirow{2}{*}{Sim~4} & 
			  $L_1$ & (1.64,0.13) & (2.34,1.14) & (1.14,0.12) \\ \cline{2-5}
			& $L_2$ & (1.65,0.13) & (2.34,1.14) & (1.14,0.12) \\ \hline
			\multirow{2}{*}{Sim~5} & 
			  $L_1$ & (1.64,0.12) & (2.44,1.09) & (1.16,0.12) \\ \cline{2-5}
			& $L_2$ & (1.66,0.12) & (2.44,1.09) & (1.16,0.12) \\ \hline
			\multirow{2}{*}{Sim~6} & 
			  $L_1$ & (1.64,0.10) & (2.54,1.65) & (1.15,0.11) \\ \cline{2-5}
			& $L_2$ & (1.65,0.11) & (2.54,1.65) & (1.15,0.11) \\ \hline
		\end{tabular}
		}
		\caption{Summary of simulation results. \label{tab-res}}
\end{subtable}
	\caption{Simulation parameters and summary. \label{tab-sim} }
\end{table}

\subsection{Quota Violation Statistics for the WS Method \label{subsec-quota-violation} }
For the simulation parameters in Table~\ref{tab-sim-param}, we obtained statistics of quota violation by the WS method. A similar study can also be carried out for the
JD method. We chose to focus only on the WS method since from Section~\ref{subsec-d2ideal} we see that the allocations obtained from the WS method are closer to the
optimal than the allocations obtained from the JD method. 

Each simulation consists of $N=1000$ iterations. Let $\mathbf{T}=(T_1,\ldots,T_n)$ be the vector of resting orders and $S$ be the size of the incoming counter-party order
in a particular iteration. Let $T=T_1+\cdots+T_n$. Further, suppose the corresponding order allocation vector obtained from the WS method is $\mathbf{S}=(S_1,\ldots,S_n)$. 
We say that $\mathbf{S}$ violates lower quota if there is an $i\in\{1,\ldots,n\}$ such that $S_i <\lfloor ST_i/T\rfloor$ and $\mathbf{S}$ violates upper quota
if there is an $i\in\{1,\ldots,n\}$ such that $S_i > \lceil ST_i/T\rceil$; $\mathbf{S}$ is said to violate quota if it violates either lower quota or upper quota.

A basic statistic to determine quota violation is the ratio of the number of iterations in which quota violation occurs to the total number of iterations.
Let $\lambda$ denote the percentage of iterations in which quota violation occurs. While $\lambda$ indicates the frequency of quota violation, it does not provide information
regarding the extent of quota violation.
The extent of lower quota violation is $S_i-\lfloor ST_i/T\rfloor$ and the extent of upper quota violation is $S_i - \lceil ST_i/T\rceil$. Note that the 
extent of lower quota violation is negative while the extent of the upper quota violation is positive. The extent of lower quota violation takes values in the
set $\{-S,\ldots,-1\}$ while the extent of upper quota violation takes values in the set $\{1,\ldots,S\}$. 
In a simulation consisting of $N$ iterations, let $u$ be the minimum value of the extent of lower quota violations and $v$ be the maximum value of the upper quota 
violations. Then the 
pair $(u,v)$ provides information about the total extent of quota violation. The values of $\lambda$, $u$ and $v$ for the different simulations are provided in
Table~\ref{tab-qv}.
From the table, we see that quota violation is quite frequent and the frequency increases with increase in the value of $n$. Further, quota violation is
spread on both the lower and the upper sides and the magnitude of the extent of both lower and upper quota violations increases with increase in the value of $n$. So
we see that for the data which models order matching application, the WS method violates quota quite frequently and also to a considerable extent. This makes the WS
method unsuitable for use as an order matching algorithm.

\begin{table}
\centering
	\begin{tabular}{|c|c|c|c|}
		 \cline{2-4}
		\multicolumn{1}{c|}{} & \multicolumn{1}{c|}{$\lambda$} & \multicolumn{1}{c|}{$u$} & \multicolumn{1}{c|}{$v$} \\ \hline
			Sim~1 & 62.7\% & $-10$ & $10$ \\ \hline 
			Sim~2 & 59.9\% & $-9 $ & $9 $ \\ \hline 
			Sim~3 & 78.8\% & $-20$ & $21$ \\ \hline 
			Sim~4 & 78.6\% & $-16$ & $28$ \\ \hline 
			Sim~5 & 86.8\% & $-30$ & $42$ \\ \hline 
			Sim~6 & 88.7\% & $-35$ & $34$ \\ \hline 
	\end{tabular}
	\caption{Statistics of quota violation by the WS method. In the table $\lambda$ is the frequency of quota violation, $u$ is the minimum
	value of the extent of lower quota violation, and $v$ is the maximum value of the extent of upper quota violation. \label{tab-qv}}
\end{table}

\section{Summary \label{sec-summary}}
In Table~\ref{tab-eff-comp} we compared the time complexities of the various methods. This showed that while the Ha and the Dr methods are marginally slower
than the pro-rata algorithm, the highest averages methods are substantially slower than the pro-rata method. 

In Table~\ref{tab-comp}, we summarise the comparison between the different algorithms
with respect to the other aspects that are relevant to the order matching application. Due to the positivity requirement, the Ad,~Da,~De,~HH methods are not suitable
(see Section~\ref{subsec-positivity}). The pro-rata, and the Ha and the Dr methods exhibit certain paradoxes, but this is not relevant to the order matching
scenario (see Section~\ref{subsec-paradox}). 

Among all possible order matching methods, the Ha method is distance optimal in the sense that allocations
produced by this method have the minimum $L_1$ and $L_2$ distances to the ideal allocation. So in a precise sense, the Ha method achieves the best possible proportionality. 
From the table, it may be observed that the Dr method possesses all the properties of the Ha method other than being distance optimal. So for the order matching problem,
there is no reason to prefer the Dr method over the Ha method.

Both the JD and the WS methods are not distance optimal and neither do they respect the quota rule. Simulation results in Section~\ref{subsec-d2ideal} show that allocations
produced by the WS method is closer to the ideal than those produced by the JD method. The simulation results of quota violation for the WS method in 
Section~\ref{subsec-quota-violation} show that the WS method violates quota quite frequently (for data generated for order matching applications). The quota violation,
not being distance optimal to the ideal allocation, and the substantial slowdown compared to the pro-rata method make both the JD and the WS methods unsuitable for order matching
applications.
\begin{table}
\centering
	\begin{tabular}{|l|c|c|c|c|c|c|c|c|c|}
		\cline{2-10}
		\multicolumn{1}{c|}{} & pro-rata & Ha & Dr & JD & WS & Ad & Da & De & HH \\ \hline
		positivity? & no & no & no & no & no & yes & yes & yes & yes \\ \hline
		paradox?    & yes & yes & yes & no & no & no & no & no & no \\ \hline
		distance optimal? & no & yes & no & no & no & no & no & no & no \\ \hline
		quota violation? & no & no & no & yes & yes & yes & yes & yes & yes \\ \hline
	\end{tabular}
	\caption{Comparison of various aspects of the different methods. \label{tab-comp} }
\end{table}

\section{Conclusion \label{sec-conclu}}
We have comprehensively investigated the possible application of various proportional representation methods as substitutes for the pro-rata order matching algorithm 
in stock exchanges. 
From our analysis, it follows that the method due to Hamilton is the most suitable candidate for the purpose. Stock exchanges may seriously consider
adopting Hamilton's method as a superior alternative to the pro-rata order matching algorithm.

\section*{Acknowledgement} We are grateful to the editor and the reviewers for providing comments which have helped in improving the paper.


\begin{thebibliography}{ACJT16}

\bibitem[ACJT16]{lim-bk2016}
Fr\'{e}d\'{e}ric Abergel, Anirban Chakraborti, Aymen Jedidi, and Ioane~Muni Toke.
\newblock {\em Limit Order Book}.
\newblock Cambridge University Press, 2016.

\bibitem[AHU78]{AHU78}
Alfred Aho, John~E. Hopcroft, and Jeffrey~D. Ullman.
\newblock {\em Design and analysis of computer algorithms, first edition}.
\newblock Addison-Wesley, 1978.

\bibitem[Bir76]{Bi76}
G.~Birkhoff.
\newblock House monotone apportionment schemes.
\newblock {\em Proceedings of the National Academy of Sciences, U.S.A.},
  73:684--86, 1976.

\bibitem[BY01]{BY01}
Michel~L. Balinski and Peyton~H. Young.
\newblock {\em Fair Representation: Meeting the Ideal of One Man, One Vote (2nd
  ed.)}.
\newblock Brookings Institution Press, 2001.

\bibitem[{Chi}]{CMEMatchingAlgorithms}
{Chicago Mercantile Exchange}.
\newblock {Supported Matching Algorithms, Clients Systems Wiki}.
\newblock
  \url{https://www.cmegroup.com/confluence/display/EPICSANDBOX/Supported+Matching+Algorithms},
  accessed on 23 February, 2023.

\bibitem[CS20]{CS2020}
Satya~R. Chakravarty and Palash Sarkar.
\newblock {\em An introduction to algorithmic finance, algorithmic trading and
  blockchain}.
\newblock Emerald Group Publishing, 2020.

\bibitem[FSS20]{FSS20}
Jaros\l{}aw Flis, Wojciech S\l{}omczy\'{n}ski, and Dariusz Stolicki.
\newblock Pot and ladle: a formula for estimating the distribution of seats
  under the {J}efferson–{D}'hondt method.
\newblock {\em Public Choice}, 182:201--227, 2020.

\bibitem[GF17]{GF17}
Josh Goldenberg and Stephen~D Fisher.
\newblock The {Sainte-Lagu\"{e}} index of disproportionality and {D}alton's
  principle of transfers.
\newblock {\em Party Politics}, 25(2), 2017.
\newblock \url{https://doi.org/10.1177/1354068817703020}.

\bibitem[Her22]{He22}
Gil Hersch.
\newblock Procedural fairness in exchange matching systems.
\newblock {\em Journal of Business Ethics}, 2022.
\newblock \url{https://doi.org/10.1007/s10551-022-05315-7}.

\bibitem[HM08]{HM08}
Carmen Herrero and Ricardo Mart\'{i}nez.
\newblock Balanced allocation methods for claims problems with
  indivisibilities.
\newblock {\em Social Choice and Welfare}, 30:603--617, 2008.

\bibitem[HPS18]{Herron2018}
Erik~S. Herron, Robert~J. Pekkanen, and Matthew~S. Shugart.
\newblock {\em The Oxford handbook of electoral systems}.
\newblock Oxford University Press, 2018.

\bibitem[Med19]{Me19}
Juraj Medzihorsky.
\newblock Rethinking the {D'Hondt} method.
\newblock {\em Political Research Exchange}, 1:1:1--15, 2019.
\newblock \url{DOI: 10.1080/2474736X.2019.1625712}.

\bibitem[Nor04]{Norris2004}
Pippa Norris.
\newblock {\em Electoral Engineering: Voting Rules and Political Behavior}.
\newblock Cambridge University Press, Mar 1, 2004 2004.

\bibitem[Pre11]{Pr11}
Tobias Preis.
\newblock {\em Price-Time Priority and Pro Rata Matching in an Order Book Model
  of Financial Markets}, pages 65--72.
\newblock Springer, 2011.
\newblock \url{https://doi.org/10.1007/978-88-470-1766-5_5}.

\end{thebibliography}

\end{document}